\documentclass[prb,amsmath,amssymb,twocolumn]{revtex4}
\usepackage{graphicx}
\usepackage{epsfig}
\usepackage{bm}

\begin{document}
\title{Magnetic field dependence of valley splitting in realistic Si/SiGe quantum wells} 
\author{Mark Friesen}
\author{M. A. Eriksson} 
\author{S. N. Coppersmith} 
\affiliation{Department of Physics, University of
Wisconsin, Madison, WI 53706} 

\begin{abstract}
The authors investigate the magnetic field dependence of the energy splitting between low-lying
valley states for electrons in a Si/SiGe quantum well tilted with 
respect to the crystallographic axis.  The presence of atomic steps at the quantum well
interface may explain the 
unexpected, strong suppression of the valley splitting observed in recent experiments.  
The authors find that the suppression is caused by an interference effect associated with 
multiple steps, and that the magnetic 
field dependence arises from the lateral confinement of the electronic wave function.
Using numerical simulations, the authors clarify the role of step disorder, obtaining 
quantitative agreement with the experiments.
\end{abstract}


\maketitle

Qubits in silicon are leading candidates for scalable quantum computing,
owing to their favorable and well studied materials properties.\cite{kane98,friesen03}  
Indeed, because of its prominence
in the electronics industry, silicon may be the best understood semiconducting 
material.  However, as devices continue to shrink in size, approaching the quantum 
regime, important questions arise.  Unlike direct gap semiconductors,
the conduction band structure in silicon possesses six symmetric minima or ``valleys" that 
are not at the
Brillouin zone center.  Consequently, the minima are degenerate, and must be described 
by a valley index which competes with the spin index as a relevant quantum number in
the qubit Hilbert space.\cite{eriksson04}
Therefore, to construct spin qubits in silicon, it is necessary to lift all valley degeneracy.

A silicon quantum well grown on the $[001]$ surface of strain-relaxed silicon-germanium
is under tensile strain, causing the four lateral valleys to rise significantly
in energy.\cite{herring56}  
At low temperatures, only the two low-lying valleys are populated.
The remaining two-fold degeneracy can be removed by the sharp 
confinement potential of the quantum well interface.\cite{AFS}  Theoretical estimates
suggest that the resulting valley splitting can be of the order of 
$1\,\text{meV}\approx 12$~K,\cite{boykin04} which is sufficiently large for quantum 
computing.  However, recent experiments in SiGe \cite{weitz96,koester97,lai04,goswami04} 
measure a valley splitting much smaller than the theoretical prediction.  There is 
currently no explanation for this discrepency.\cite{khrapai03}  Indeed, a
prevalent theory \cite{ohkawa77} predicts an enhancement of the 
valley splitting in a magnetic field that is different from the experimental observations. 

\begin{figure}[t]
\begin{center}
\includegraphics[width=2.1in]{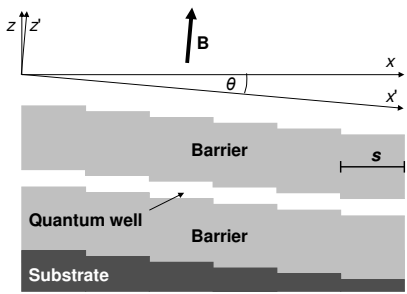}
\caption{
Quantum well step geometry, with 
crystalographic axes $(x,y,z)$ and rotated axes $(x',y',z')$, where $y=y'$.
\label{fig:oscillations}}
\end{center}
\end{figure}

In this letter, we describe a single-electron valley splitting theory for silicon quantum 
wells grown on a vicinal substrate, building upon an initial suggestion by 
Ando.\cite{ando79}  Such miscuts are often
incorporated into Si/SiGe heterostructures to ensure uniform growth surfaces and to
avoid step bunching.  The 
resulting quantum well, obtained by conformal epitaxial deposition, is 
misaligned with respect to the crystallographic $z$ axis, as shown in 
Fig.~\ref{fig:oscillations}.  We now describe the effective mass theory and explain
how the presence of interfacial atomic steps suppresses valley splitting.

For silicon strained in the $[001]$ direction, the effective mass wavefunction \cite{kohn} can 
be written as a sum of contributions from the two $z$ valleys:\cite{friesenunp}
\begin{equation}
\Psi({\bm r}) =  \frac{1}{\sqrt{2}}
\left[ e^{ik_0z}u_{k_0}({\bm r})+ e^{-ik_0z+i\phi}u_{-k_0}({\bm r}) \right]
F({\bm r}) . \label{eq:kohn}
\end{equation}
where the terms inside brackets are Bloch functions.  
The phase angle $\phi$ is determined by the position of the quantum well 
interface.\cite{friesenunp}  The two orthogonal valley states correspond to a phase difference
of $\Delta \phi =\pi$, although the absolute value of $\phi$ is unimportant in the present
discussion.  It is crucial to note that while the valley minima occur along the $z$ axis at 
$\pm k_0\hat{\bm z}$, the quantum well normal is tilted away from $\hat{\bm z}$.  For a 
slowly varying confinement potential, the two valleys have the same envelope function, and 
are essentially independent.  However, sharp
variations in the potential cause the effective mass approximation to break down.  Examples 
include the central cell potential near a shallow donor,\cite{friesenprl} and the 
sharp band offsets at the interface of a quantum well.\cite{friesenunp}
Leading order corrections to the effective mass theory \cite{friesenunp} give the valley 
splitting
\begin{equation}
E_v=2\left| \int dr^3 \, e^{-i 2k_0 z} |F({\bm r})|^2 V_v({\bm r}) \right| ,
\label{eq:Dv}
\end{equation}
where the coupling potential $V_v({\bm r})$ decays a few Angst\"{o}ms from the interface.
Because this decay length is so small, compared to effective mass length scales,
it can be represented as a $\delta$ function:
\begin{equation}
V_v({\bm r}) = v_v \delta (z-z_i) ,
\label{eq:Vi}
\end{equation}
where $z_i(x)$ is the position of the interface along the $x$ axis.
The coupling parameter $v_v$ contains atomic scale information that must be obtained from 
\textit{ab initio} theories such as tight-binding theory,\cite{friesenunp} or from experiments.
The phase factor in Eq.~(\ref{eq:Dv}) reflects the valley separation of $2k_0$, and 
plays a crucial role in the theory by introducing interference effects, as we shall see.
Note that large modulation doping fields usually confine the wave function to one side of the 
quantum well, so only one interface potential has been included in Eq.~(\ref{eq:Vi}). 

We can now explain the suppression of the valley splitting.
In a tilted quantum well, the interface position $z_i(x)$ describes the atomic steps, 
as shown in Fig.~\ref{fig:oscillations}.  The $\delta$ function in 
Eq.~(\ref{eq:Vi}) reduces the $E_v$ integral to a sum over steps with phase angles 
differing by $2k_0b\simeq 0.85 \pi$.  Here, $k_0 \simeq 0.85 (\pi /2b)$, and 
$b\simeq 1.358$~\AA~is the atomic step height.  
Thus, the phases from consecutive steps interfere almost fully destructively.
For a delocalized electron, the wavefunction 
extends over an infinite number of steps, resulting in the complete suppression of $E_v$.  
In the presence of a magnetic field, the electronic wave function is confined to a 
finite number of steps, leading to a non-vanishing valley splitting.

We can estimate the valley splitting at low magnetic fields from Eq.~(\ref{eq:Dv}), 
assuming the idealized (but
unrealistic) miscut geometry shown in Fig.~\ref{fig:oscillations}, with uniform steps
of equal width.  A high-field tight-binding theory was presented in 
Ref.~\onlinecite{leepreprint} for the same geometry.  A good approximation for 
the  wave function of an electron in a perpendicular magnetic field and a tilted 
quantum well is given by the well-known solution for a flat quantum well,\cite{Daviesbook} 
in the rotated basis $(x',y',z')$ shown in Fig.~\ref{fig:oscillations}.
For the lowest ($n=0$) eigenstate corresponding to the symmetric gauge,  
${\bm A}=(-y',x',0)B/2$, we obtain \cite{Daviesbook} $F({\bm r}')=F_{xy}(x',y')F_z(z')$, 
where $F_{xy}(x',y')= e^{-(x'^2+y'^2)/4l_B^2}/\sqrt{2\pi l_B^2}$ and 
$F_z(z')$ is the subband envelope.  Here, the electronic  wave function is confined over the 
magnetic length scale $l_B=\sqrt{\hbar/|eB|}$.  For large $l_B$, we can ignore the
discreteness of the atomic steps in $z_i(x)$.  Performing the integral in 
Eq.~(\ref{eq:Dv}), we obtain 
$E_v\simeq 2v_vF_z^2(0)e^{-2(k_0l_B\theta )^2}$, where $\theta \ll 2\pi$ is the
miscut angle.  Thus, for uniform steps, 
the valley splitting is suppressed exponentially in $B$, 
as confirmed by the numerical analysis described below.

\begin{figure}[t]
\begin{center}
\includegraphics[width=3.2in]{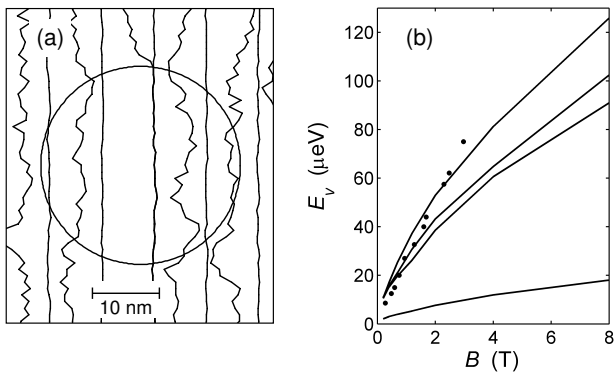}
\caption{
Realistic valley splitting simulations.
(a) A randomly generated step edge profile, with 
alternating smooth and rough step edges.   
The circle shows the  wave function radius $l_B$ for
$B=3$~T, centered at the position of the valley splitting maximum.
(b) Simulation results for the valley splitting (solid lines).  Experimental 
data points (circles) are taken from Ref.~\onlinecite{goswami04}.  
The top three curves assume the same ``large" amplitude of step fluctuations,
with different step bunching parameters.
From top to bottom: (i) strong bunching, (ii) weak bunching, 
(iii) no bunching, (iv) fluctuation amplitude reduced by a factor of 5.
Note that the top curve corresponds to the disorder realization in (a).
\label{fig:stepSTM}}
\end{center}
\end{figure}

The previous results are strongly affected by even a little disorder.  At low
fields, disorder can enhance the valley splitting by many orders of magnitude, showing that 
the exponential suppression occurs because of a delicate cancellation of the phase 
terms in Eq.~(\ref{eq:Dv}).  
To investigate the effects of disorder, we have performed 
numerical simulations of the valley splitting.  We introduce adjustable  
parameters describing the amplitude of normally-distributed fluctuations, or wiggles,
of the step edges.  The fluctuation model incorporates alternating smooth 
and rough step edges, as consistent with the experimental data.\cite{lagally}
Typically in our simulations, the amplitude of the rough step fluctuations is set at its 
characteristic 
``large" value $s$, corresponding to the average step separation.
We also introduce a step-bunching parameter, which allows neighboring step edges to 
overlap.  Such bunching is known to occur at silicon 
growth interfaces, particularly under the influence of strain.\cite{tersoff}
A representative step profile is shown in Fig.~\ref{fig:stepSTM}(a).  To make contact 
with the data of Ref.~\onlinecite{goswami04}, we consider a $2^\circ$ miscut.  

For a given, randomly generated step profile, we evaluate Eq.~(\ref{eq:Dv}) numerically.
The resulting valley splitting now depends on the electron position,
since disorder breaks the translational symmetry.  In the symmetric
gauge, the unperturbed magnetic  wave function can be centered, degenerately, at different  
points in space.  
By computing the valley splitting for  wave functions centered at each of these points,
we obtain a valley splitting landscape, $E_v({\bm r}')$,
including peaks and valleys.  Electrons are attracted to the valley splitting peaks 
in order to minimize their total energy.  We observe that peaks always occur near
broad step fluctuations, or ``plateaus,"  and that bunched steps tend to 
produce the strongest valley splitting peaks.  By tracking
the dominant peak in a given cell, we can determine $E_v(B)$ for a fixed
step profile.  In Eq.~(\ref{eq:Vi}) we use 
$v_v\simeq 1.2\times 10^{-11}\,\text{eV}\cdot\text{m}$, as obtained in 
Ref.~\onlinecite{friesenunp}, which corresponds to a 
Si$_{0.7}$Ge$_{0.3}$/Si/Si$_{0.7}$Ge$_{0.3}$ quantum well.

Some typical simulation results for $E_v(B)$ are shown in Fig.~\ref{fig:stepSTM}(b).  
By setting the step fluctuations to their characteristic ``large" value, as described
above, we obtain reasonable
agreement with the experimental data of Ref.~\onlinecite{goswami04} for
any value of the step bunching parameter.  
By fine tuning the step-bunching parameter, we obtain the best quantitative
agreement when the average bunching length is in the range of 10-$20s$.

In the intermediate field range $0.3<B<3$~T, the experimental data points in 
Fig.~\ref{fig:stepSTM} are strikingly linear.\cite{goswami04}
The corresponding simulation results exhibit a slightly sublinear field dependence.
More generally, we find that the precise shape of the $E_v(B)$ 
curves depends on the particular disorder realization and the fluctuation model.  It is
very likely that other disorder 
models, not considered here, could produce different curve shapes, including the linear
behavior of the data.  One idealized model giving a linear valley splitting is the 
``plateau" model, corresponding to wide, localized, flat plateaus surrounded by 
relatively uniform steps.  The valley splitting contribution from the ordered steps surrounding 
the plateau is strongly suppressed, due to the cancellation effects described above, so 
that the plateaus dominate the valley splitting.
The scaling theory for an isolated plateau is obtained by introducing the concept of
an ``excess area" $A$ relative to a perfectly uniform step configuration.  For a small
magnetic field, the electronic
 wave function is spread out over the large area $2\pi l_B^2$.
The resulting valley splitting in this model is
\begin{equation}
E_v \simeq v_vF_z^2(0)A/\pi l_B^2 ,
\label{eq:linear}
\end{equation}
which is linear in $B$.
We can obtain an experimental estimate for $A$ from the data of 
Ref.~\onlinecite{goswami04}.  Using theoretical estimates for $v_v$
and $F_z(0)$,\cite{friesenunp} we find that $A\simeq 18 s^2$, where $s=3.9$~nm is the 
average step separation for a $2^\circ$ miscut.  This result in general agreement with our 
simulations.  For a round plateau, this corresponds to a diameter of about $5s$.

The plateau scaling model must break down for very small 
fields (large $l_B$), when the  wave function encloses additional plateaus. 
At high fields, the scaling also breaks down when
the  wave function is confined to a single step,
saturating at the theoretical upper bound of Ref.~\onlinecite{boykin04}.  
For a $2^\circ$ miscut,
this crossover occurs at about 25~T.  Thus, the scaling expression in Eq.~(\ref{eq:linear}),
and possibly the linear experimental data, may correspond to crossover 
behavior.  Finally, we note that the magnetic field dependence of the valley splitting 
should be first order in the valley coupling parameter
$v_v$, consistent with Eq.~(\ref{eq:linear}), since it involves breaking the 
positional degeneracy of the magnetic (Landau) eigenfunctions.

Valley states can be detrimental for spin-based quantum computing.\cite{koiller02} 
Large valley splittings are needed to avoid 
thermal excitations outside the spin-1/2 qubit Hilbert space associated with
the valley ground state.\cite{friesen03}  To accomplish this 
in an experimental setting, the present results suggest that we should use 
substrates without miscuts, although it is usually difficult to eliminate large wavelength
roughness in conventional, strained devices.  
An alternative approach is to develop heterostructures utilizing strain-\textit{sharing}
techniques,\cite{michele} which can, in principle, provide step-free interfaces.
Indeed, large valley splittings have be obtained in similar unstrained silicon oxide 
structures.\cite{takashina06}

Finally, we point out that a tilted magnetic field can be used to test the proposed
mechanisms for valley splitting suppression.
When the magnetic field is tilted away from the growth axis ($\hat{\bm z}'$), 
we would expect strong variations in the valley splitting, depending on the relative
angle of the field with respect to the quantum well.

In summary, we have demonstrated that experimental observations of the magnetic 
field dependence of the valley splitting are consistent with a theory in which 
interference between different atomic steps at an interface causes the valley splitting 
to decrease, as the size of the electronic  wave function increases.  The present results 
suggest that valley splitting can be 
increased by using substrates without miscuts, although it is usually difficult 
to eliminate large wavelength roughness in conventional, strained devices.  Alternatively,
valley splitting can be controlled by constraining the  wave function with electrostatic gates,
as demonstrated in Ref.~\onlinecite{goswami2}.

\begin{acknowledgments}
The authors would like to acknowledge discussions with S. Chutia, S.~Goswami, R.~Joynt, 
G.~Klimeck, C.~Tahan, and P.~von Allmen.
This work was supported by NSA and ARDA under ARO Contract 
No.~W911NF-04-1-0389 and by the National Science Foundation through the ITR 
(DMR-0325634) and EMT (CCF-0523675) programs.
\end{acknowledgments}

\end{document}